\documentclass[preprint,12pt, a4paper]{elsarticle}

\makeatletter
\def\ps@pprintTitle{%
 \let\@oddhead\@empty
 \let\@evenhead\@empty
 \let\@oddfoot\@empty
 \let\@evenfoot\@empty
 }
\makeatother

\usepackage{amssymb}
\usepackage{hyperref}
\usepackage{amsmath}
\usepackage{listings}
\usepackage{color}
\definecolor{mygreen}{rgb}{0,0.6,0}
\definecolor{mygray}{rgb}{0.5,0.5,0.5}
\definecolor{mymauve}{rgb}{0.58,0,0.82}
\definecolor{greybg}{rgb}{0.9,0.9,0.9}
\journal{SoftwareX}

\begin{document}
\renewcommand{\labelenumii}{\arabic{enumi}.\arabic{enumii}}

\begin{frontmatter}

\title{QHyper: an integration library for hybrid \\quantum-classical optimization}


\author[label1,label2]{Tomasz Lamża}
\ead{tomasz.lamza@cyfronet.pl}
\author[label1,label2]{Justyna Zawalska}
\ead{justyna.zawalska@cyfronet.pl}
\author[label1,label2]{Kacper Jurek}
\ead{kacper.jurek@cyfronet.pl}
\author[label2]{Mariusz Sterzel}
\author[label1,label2]{Katarzyna Rycerz}
\ead{kzajac@agh.edu.pl}
\address[label1]{AGH University of Krakow, Institute of Computer Science, al. Mickiewicza 30, 30-059 Krakow, Poland}
\address[label2]{Academic Computer Center Cyfronet AGH, Nawojki 11 Street, 30-950 Krakow, Poland}

\begin{abstract}
We propose the QHyper library, which is aimed at researchers working on computational experiments with a variety of quantum combinatorial optimization solvers. The library offers a simple and extensible interface for formulating combinatorial optimization problems, selecting and running solvers, and optimizing hyperparameters. The supported solver set includes variational gate-based algorithms, quantum annealers, and classical solutions. The solvers can be combined with provided local and global (hyper)optimizers. The main features of the library are its extensibility on different levels of use as well as a straightforward and flexible experiment configuration format presented in the paper.\\

\end{abstract}


\end{frontmatter}


\section*{Metadata}
\label{}

\begin{table}[!h]
\begin{tabular}{|l|p{4.4cm}|p{8.6cm}|}
\hline
C1 & Current code version & 0.2.2 \\
\hline
C2 & Permanent link to code/repository used for this code version & \url{https://github.com/qc-lab/QHyper/releases/tag/v0.2.2} \\
\hline
C3  & Permanent link to Reproducible Capsule & \url{https://codeocean.com/capsule/1259194/tree}\\
\hline
C4 & Legal Code License   & MIT License \\
\hline
C5 & Code versioning system used & git \\
\hline
C6 & Software code languages, tools, and services used & Python, PennyLane, D-Wave \\
\hline
C7 & Compilation requirements, operating environments \& dependencies & Linux, OS X \\
\hline
C8 & If available Link to developer documentation/manual & \url{https://qhyper.readthedocs.io/en/latest/} \\
\hline
\end{tabular}
\caption{Code metadata}
\label{codeMetadata} 
\end{table}



\section{Motivation and significance}
\label{sec:motivation}


\par 
Currently, there are many examples of research going towards achieving the so-called quantum advantage, i.e. a situation when a quantum computer can solve problems that are beyond the reach of a classical one. One of the domains that could benefit from this approach is combinatorial optimization due to its classical exponential complexity. Therefore, determining whether hybrid or  purely quantum algorithms could solve optimization problems more efficiently than state-of-the-art classical solvers would bring a major advancement and still remains an open question~\cite{abbas_quantum_2023}.\\ 

The most common  quantum  approaches  supporting  combinatorial optimization are based on two different computational models: adiabatic quantum computing~\cite{mcgeoch_adiabatic_2014} and  variational algorithms designed for universal, gate-based quantum devices~\cite{cerezo_variational_2021}. This support is often realized by developing optimization solvers implemented in multiple software libraries. Additionally, for researchers working with those solutions, it is often necessary to evaluate the quality of a quantum device's output by comparing it to the results obtained from established classical optimization packages. Therefore, there is a lack of a common API that could be used to connect all  solvers necessary for the researcher's convenience in one place. \\  

In this paper, we addresses this need by introducing  the QHyper library, which  provides the interface for specifying combinatorial optimization problems, choosing solvers (including those based on variational algorithms as well as quantum annealers and reference  classical solutions), and dealing with problem hyperparameters. All necessary properties of an optimisation experiment can be expressed using a simple configuration file. \\

Due to its extensible interface and easy experiment configuration, QHyper is suitable for various user needs. Firstly, it can be used as a standalone tool by users who want to experiment with solving combinatorial optimization problems using quantum or hybrid quantum-classical methods. The user only implements or imports a combinatorial problem and configures the desired type of solver. This allows, e.g., to investigate which versions of problem representations (method of variables encoding, method of handling inequality constraints, etc.) are best suited to achieving good results with quantum or hybrid algorithms.\\

Additionally, this tool is suitable for assessing new solvers. For example, the realm of improvements to the commonly used  Variational Quantum Algorithms (VQA) such as the Variational Quantum Eigensolver~\cite{peruzzo_variational_2014} and the Quantum Approximate Optimization Algorithm (QAOA)~\cite{farhi_quantum_2022} is constantly growing. Having needed solvers implemented in one place makes it easy to compare their performance, as was shown in~\cite{mikyska_software_2023}.\\

QHyper is also aimed at users who want to test new (hyper)parameter optimization algorithms for existing solvers. This is particularly important in cases where hyperparameters are necessary to properly express the problem formulation (e.g. penalty weights needed to combine the cost function and constraints), or where fine-tuning parameters is a significant aspect of the solver's operation, as in VQAs.\\



Although there are several examples of hybrid quantum-classical and quantum computing optimization software, most of them specialize exclusively in algorithms for gate-based solvers (e.g., Qiskit Optimizers \cite{matthew_treinish_qiskitqiskit_2024}, OpenQAOA \cite{sharma_openqaoa_2022}) or adiabatic quantum computing (e.g., D-Wave Ocean \cite{noauthor_d-wave_nodate}). To the best of our knowledge, there are only a few software tools unifying solvers for both of these techniques, namely QPLEX~\cite{giraldo_qplex_2023}, QC Ware Forge~\cite{noauthor_qc_nodate}, and Qibo~\cite{carrazza_open-source_2023}. However, none of the existing tools currently offer an interface to accessing/defining (hyper)optimizers or enable specifying experiment properties using a simple configuration file. There are also other differences. \\

In QPLEX, problems are formulated using DOcplex mathematical programming model. To translate the problem description into a Quadratic Unconstrained Binary Optimization (QUBO)~\cite{glover_quantum_2022} formulation QPLEX uses methods provided by IBM Qiskit. Conversely, in QHyper, users have the flexibility to express problem models in a variety of formats, including custom ones. While QHyper employs a similar approach to problem representation as QC Ware Forge \cite{noauthor_qc_nodate}, it differs from Forge in that it is not limited to integer-coefficient functions. Additionally, the optimization method used in Forge has a significant number of parameters in its signature. This can be a challenge for configuring the experiments, especially given the fact that this software is proprietary. In Qibo \cite{efthymiou_qibo_2022}, optimization problems should be defined in the form of a Hamiltonian. In contrast, QHyper offers users the flexibility to define optimization problems using various methods (including expressing them in readable formats like Sympy expressions). Additionally, at present, Qibo does not integrate with D-Wave. \\


\section{Software description}
As described in Section~\ref{sec:motivation}, QHyper is a   modular software tailored to the specific needs of different combinatorial optimization tasks. In this section, we present details about its main components and functionality.

\subsection{Software architecture}
\begin{figure}[!h]
    \centering
    \includegraphics[scale=0.45]{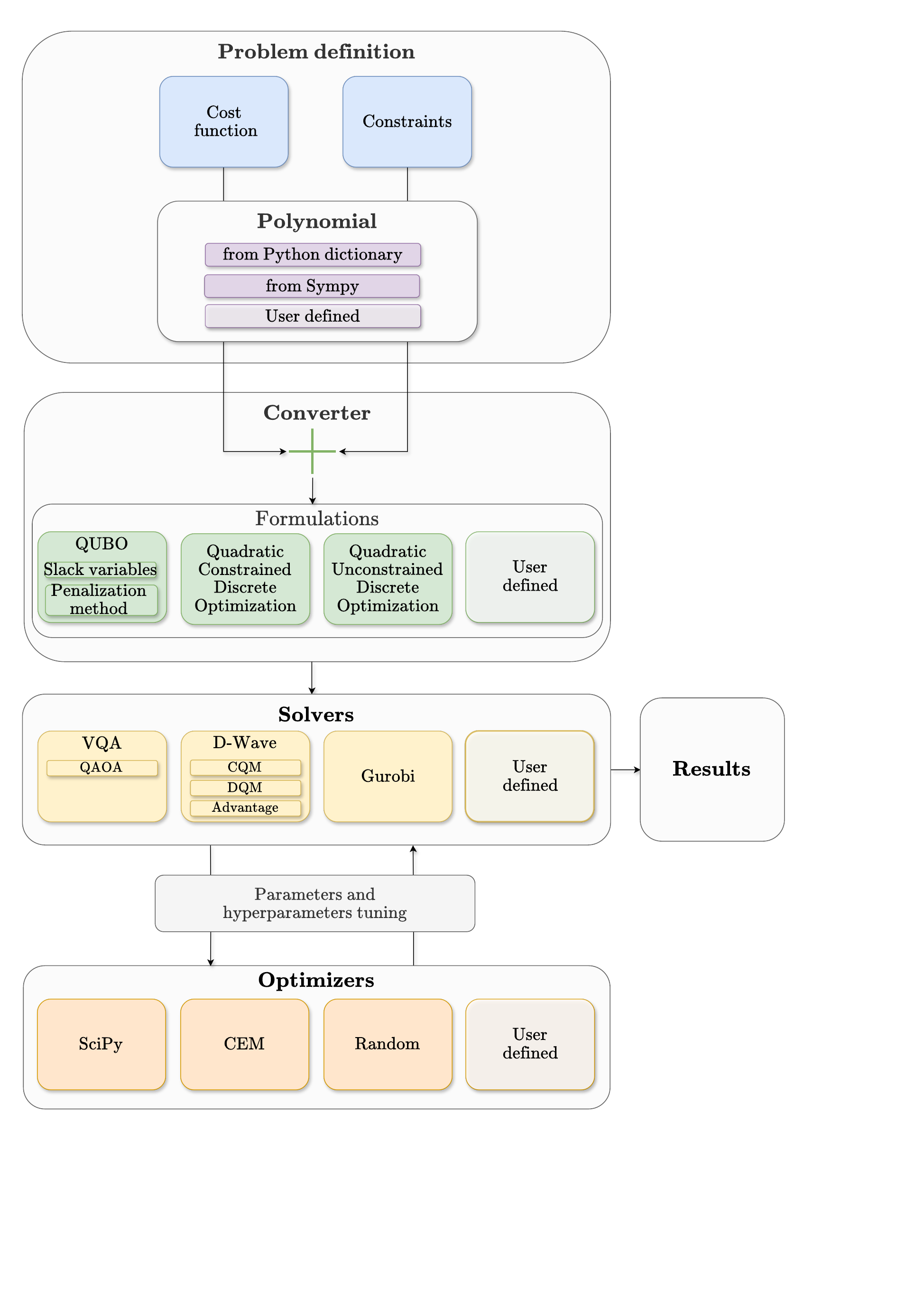}
    \caption{Architecture of QHyper}
    \label{fig:QHyperarch}
\end{figure}

The architecture of the system is shown in Fig.~\ref{fig:QHyperarch}. 
The high-level scheme for utilizing the software revolves around the following key elements: describing  the problem, running a selected solver with accurate properties, and optionally selecting a (hyper)optimizer to enhance the solver's performance. For a seamless data transition between the problem definition and the solver input, it is necessary to convert the provided problem description ({\em Problem definition} module) into a format that the chosen solver can understand. This is done by combining the problem's {\em cost function} and its {\em constraints} into the appropriate formulation (this is indicated as a plus sign in the {\em Converter} module). Also, depending on the solver type ({\em Solvers} module), the user may need to define additional optimizers for the parameters or hyperparameters used within the solver ({\em Optimizers} module). After optimization, the solver returns the probabilities of specific results and, if applicable, the optimized (hyper)parameters.  Below, we provide a detailed description of the main system components.

\subsubsection{Problem definition}
Combinatorial optimization problems usually consist of a cost function and constraints. While QHyper can handle unconstrained problems, our focus is on addressing problems with constraints because they are more common in real-world examples. QHyper already implements a few popular problems such as the Knapsack Problem~\cite{lucas_ising_2014} or the Traveling Salesman Problem~\cite{lucas_ising_2014}, as well as dedicated examples from diverse fields of science ~\cite{krzhizhanovskaya_foundations_2020,wierzbinski_community_2023}. Furthermore, users can effortlessly add new problems. \\

\par In order to solve problems using various (often third-party) solvers, it was necessary to establish a common representation of the problem that can be easily translated into the format expected by the given solver. To achieve this standardization, we created the Polynomial format which is utilized to describe an objective function, as well as the left-hand and right-hand sides of every constraint. \\

\par A Polynomial is comprised of a dictionary where the keys are tuples containing variables, and the values represent their coefficients. Using dictionaries allows for efficient arithmetic operations on Polynomials, simplification of terms, and extraction of relevant information such as constants, degree, and variables. The core functionality includes addition, subtraction, multiplication, exponentiation, and negation. This representation can store higher-order polynomials. The creation of a Polynomial can be done manually by providing a dictionary or by translating it from SymPy syntax. Additionally, users can implement the translation into Polynomial from their own data source (see components of {\em Polynomial} box in Fig.~\ref{fig:QHyperarch} ).

\subsubsection{Converter}
Different solvers require different problem formulations. QHyper currently supports the QUBO, Quadratic Constrained Discrete Optimization\footnote{\url{https://docs.ocean.dwavesys.com/en/latest/concepts/cqm.html}} and Quadratic Unconstrained Discrete Optimization formulations\footnote{\url{https://docs.ocean.dwavesys.com/en/latest/concepts/dqm.html}}. Additionally, users can integrate custom ones if required (see {\em Formulation} module in Fig.~\ref{fig:QHyperarch}). To facilitate this feature, we have created the Converter, which contains multiple methods for obtaining a solver-specific form of the problem.\\

\par For instance, the QUBO representation of the constrained problem is a weighted sum of the cost function and actual constraints.  These weights are referred to as penalty weights (or Lagrangian multipliers)  as they penalize results that do not meet the restrictions. Also, inequality constraints have to be converted into equality constraints, and this can be accomplished in QHyper in several ways, such as adding slack variables~\cite{glover_quantum_2022} or using the unbalanced penalization method~\cite{montanez-barrera_unbalanced_2024}. 

\subsubsection{Solvers}
\par Solvers can be divided into two categories: internal and third-party. Internal solvers use the algorithms implemented within QHyper. Currently, these include solvers based on the use of gate-based quantum algorithms, specifically the VQAs. These consist of the QAOA and its two variants~\cite{mikyska_software_2023}, which are implemented using the PennyLane library~\cite{bergholm_pennylane_2018}.\\

\par On the other hand, third-party solvers are interfaces to existing software packages or frameworks, providing users with a convenient way to access and use popular optimization tools. Such solvers include D-Wave\footnote{\url{https://www.dwavesys.com/}} that provides hybrid quantum-classical and quantum annealers and Gurobi\footnote{\url{http://www.gurobi.com}} that serves as a reference method that uses purely classical optimization. To use these functionalities it is necessary for a user to separately obtain required access tokens in accordance with the proper license.\\

\par The QHyper platform enables users to add new algorithms to the solver set (see {\em Solver} module in Fig.~\ref{fig:QHyperarch}). This flexibility allows users to enhance the platform's functionality and adapt it to their needs.\\

\subsubsection{Optimizers}

\par Often solvers need to be supported by additional classical optimizers.
Optimizers in QHyper can be split into two groups: local and global. Local optimizers use a variety of algorithms based on gradient descent~\cite{noauthor_numerical_2006} to find the best result. Our software  currently supports optimizers from SciPy\footnote{\url{https://docs.scipy.org/}} and PennyLane\footnote{\url{https://docs.pennylane.ai/}}. Their typical use case is  {\tt optimizer} in QAOA (see Listing \ref{lst:config_localoptim} explained in detail in Section~\ref{sec:configexa}), for variational parameters of an ansatz (a quantum circuit) related to the problem. 
Global optimizers usually use some heuristics (such as Monte Carlo methods~\cite{rubinstein_cross_2004}) to search the whole solution space.  
Their typical use case is   {\tt hyper\_optimizer} for searching for initial variational parameters for QAOA or for searching the best hyperparameters such as weights (penalties, Lagrangian multipliers) in the QUBO formulation (see Listing~\ref{lst:config_hyperoptim} and the corresponding explanation in Section~\ref{sec:configexa}). In QHyper there are implemented a few different global optimizers: random search, grid search and  Cross Entropy Method (CEM)~\cite{roch_cross_2020,paszynski_cross_2021} (see {\em Optimizers} module in Fig.~\ref{fig:QHyperarch}). These two typical use cases are often combined, e.g., the user can use the random or CEM hyperoptimizer that actually launches many instances of QAOA with a local gradient optimizer (see the example Listing~\ref{lst:config_alloptim} further described in Section~\ref{sec:configexa}).

\section{Illustrative examples }
\subsection{Sample solver configurations}
In this section, we provide examples of different solver configurations for the sample Knapsack Problem with three items shown in Listing~\ref{lst:config_problem}. The goal is to put chosen items in the knapsack to achieve maximal cost  with total weight not exceeding {\tt max\_weight}.  The {\tt items\_weights} and {\tt item\_values} sections specify the weight and cost of each item, respectively. 
\label{sec:configexa}
\begin{lstlisting}[caption={Sample configuration for the  {\tt Knapsack Problem} with three items. 
},label={lst:config_problem}]
problem:
  type: knapsack
  max_weight: 2
  items_weights: [1, 1, 1]
  items_values: [2, 2, 1]
\end{lstlisting}
To execute solvers configured in Listings \ref{lst:config_localoptim}-\ref{lst:config_alloptim}, QHyper's converter automatically creates the QUBO objective function of the problem (\ref{kqubo})
\begin{equation}
\label{kqubo}
    f(\boldsymbol{x}, \boldsymbol{y}) = - \alpha_0 \underbrace{\sum_{i = 1}^N c_i x_i}_{\text{cost function}} + \alpha_1 \underbrace{(1 - \sum_{i=1}^W y_i)^2}_{\text{encoding constraint}} + \alpha_2 \underbrace{(\sum_{i=1}^W iy_i - \sum_{i=1}^N w_ix_i)^2}_{\text{weight constraint}}, 
\end{equation}
where $N=3$ is the number of items available, $W$={\tt max\_weight} is the maximum weight of the knapsack, $c_i$ and $w_i$ are the values and weights specified in {\tt items} list of the configuration. The goal is to optimize $\boldsymbol{x} = [x_i]_N$ which is a Boolean vector, where $x_i = 1$ if and only if the item $i$ was selected to be inserted into the knapsack. $\boldsymbol{y} = [y_i]_W$ is a one-hot vector where $y_i = 1$ if and only if the weight of the knapsack is equal to $i$; $\alpha_j$ are penalty weights (hyperparameters of the optimized function).
\subsubsection{Typical QAOA configuration}
Listing~\ref{lst:config_localoptim} shows a typical QHyper configuration of QAOA with 5 {\tt layers} and the default local Adam gradient descent {\tt optimizer}~\cite{kingma_adam_2017} from PennyLane ({\tt qml}) with default options. Initial variational parameters optimized by the Adam method are set as {\tt angles}. The setting of {\tt hyper\_args} refers to the initial penalty weights in the optimized function of the  Knapsack Problem i.e. $\alpha_j$ in (\ref{kqubo}). In this simple configuration, they  remain unchanged during the whole optimization process as there is no hyperoptimizer defined in the configuration.

\begin{lstlisting}[caption={
QHyper configuration of QAOA with 5  {\tt layers} and  local gradient descent {\tt optimizer} (qml); {\tt angles} indicate  variational parameters searched  by specified {\tt optimizer}; {\tt hyper\_args} refer to the initial weights in the objective function of the  {\tt Knapsack Problem}. 
},label={lst:config_localoptim}]
solver:
  type: vqa
  pqc:
    type: qaoa
    layers: 5
  optimizer:
    type: qml
  params_inits:
    angles: [[0.5, 0.5, 0.5, 0.5, 0.5], [1, 1, 1, 1, 1]]
    hyper_args: [1, 2.5, 2.5]
\end{lstlisting}
\subsubsection{Sample  hyperoptimizer configuration with D-Wave Advantage system}
Listing~\ref{lst:config_hyperoptim} shows the sample configuration of hyperoptimizer. In this example, the {\tt grid} search hyperoptimizer is applied to find  proper penalties  ($\alpha_j$ in (\ref{kqubo})) of the  knapsack optimization function.
The penalties are searched within the specified {\tt bounds} with {\tt steps} defined in the configuration.  
The actual problem  is solved by  quantum annealing {\tt advantage} system offered by D-Wave (here the default advantage\_system5.4 is chosen). The machine is sampled 100 times ({\tt num\_reads} parameter).
\begin{lstlisting}[caption={Sample configuration for a grid search hyperoptimizer; the  objective function penalties ({\tt hyper\_args}) are searched within specified {\tt bounds} using  provided {\tt steps}; the objective function is solved with D-Wave Advantage},
label={lst:config_hyperoptim}]
solver:
  type: advantage
  num_reads: 100
  hyper_optimizer:
    type: grid
    steps: [0.1, 0.1, 0.1]
    bounds: [[1, 10], [1, 10], [1, 10]]
\end{lstlisting}
\subsubsection{Sample advanced configuration}
Listing~\ref{lst:config_alloptim} shows an advanced QHyper configuration using both {\tt optimizer} and {\tt hyperoptimizer}. The example provides usage of the QAOA variant called WF-QAOA ~\cite{mikyska_software_2023} with 5 {\tt layers}, local Adam {\tt optimizer} from PennyLane ({\tt qml}) with its custom  {\tt steps and stepsize} parameters. Additionally, PennyLane backend simulator was set to {\tt default.qubit}. Initial variational parameters of WF-QAOA are set in the configuration as {\tt angles}. As before, {\tt hyper\_args}  refer to the initial penalties in the objective function of the Knapsack Problem ($\alpha_j$ in (\ref{kqubo})). In contrast to Listing~\ref{lst:config_localoptim}, they are  further optimized by the global {\tt CEM } hyperoptimizer within specified {\tt hyper\_optimizer} bounds.
The parameters {\tt processes}, {\tt samples\_per\_epoch}, and {\tt epochs} are specific to the CEM method~\cite{paszynski_cross_2021}.

\begin{lstlisting}[caption={
QHyper configuration of the QAOA variant (WF-QAOA) with 5  {\tt layers} and  the local gradient descent Adam {\tt optimizer} (qml);  {\tt angles} indicate  initial variational parameters optimized by the method; {\tt hyper\_args} refer to the initial objective function penalties searched within {\tt hyper\_optimizer} {\tt bounds} by the {\tt CEM} method; {\tt processes}, {\tt samples\_per\_epoch}, and {\tt epochs} are parameters specific to the {\tt CEM} method. 
},label={lst:config_alloptim}]
solver:
  type: vqa
  pqc:
    type: wfqaoa
    layers: 5
    backend: default.qubit
  optimizer:
    type: qml
    optimizer: adam
    steps: 50
    stepsize: 0.01
  hyper_optimizer:
    type: cem
    processes: 4
    samples_per_epoch: 200
    epochs: 10
    bounds: [[1, 10], [1, 10], [1, 10]]
  params_inits:
    angles: [[0.5, 0.5, 0.5, 0.5, 0.5], [1, 1, 1, 1, 1]]
    hyper_args: [1, 2.5, 2.5]
\end{lstlisting}



\subsection{Sample snippets for solver execution and results evaluation}
The solver configurations presented in Section~\ref{sec:configexa} can be used as experiment properties for a researcher working on optimization problems.
A significant advantage of QHyper library is the possibility to define these properties in YAML format (or similar, e.g. JSON, which can be parsed to Python dict). Listing \ref{lst:solve} presents how to run the solver from one of the Listings~\ref{lst:config_localoptim}-\ref{lst:config_alloptim} with the problem defined in Listing~\ref{lst:config_problem} (both solver and problem listings are stored in {\tt configuration.yaml} file). The results are returned in a SolverResults object, which contains probabilities of results (encoded as numpy recarray). Additionally, a full history is returned to allow for further investigations and conclusions about the solver that was used.



\begin{lstlisting}[caption={Runnig the solver from the provided config.},label={lst:solve}]
import yaml
from QHyper.solvers import solver_from_config

with open(f"configuration.yaml", 'r') as stream:
    config = yaml.safe_load(stream)

solver = solver_from_config(config)
results = solver.solve()
results.probabilities

rec.array([(0, 0, 0, 0, 0, 0.00305416), (0, 0, 0, 0, 1, 0.0118166 ),
           (0, 0, 0, 1, 0, 0.05395579), (0, 0, 0, 1, 1, 0.0135544 ),
           ...
           (1, 1, 1, 0, 0, 0.01474204), (1, 1, 1, 0, 1, 0.01545693),
           (1, 1, 1, 1, 0, 0.03421446), (1, 1, 1, 1, 1, 0.02285358)],
          dtype=[('x0', '<i4'), ('x1', '<i4'), ('x2', '<i4'), ('x3', '<i4'), ('x4', '<i4'), ('probability', '<f8')])

\end{lstlisting}

QHyper also provides a few functions to obtain useful information from results, such as evaluation (i.e. displaying the value of the problem cost function for the cases where the constraints are met and 0 otherwise), sorting and limiting results, and calculating expectation value. Listing 
\ref{lst:post} demonstrates how to use these functions.

\begin{lstlisting}[caption={Further operations on results.},label={lst:post}]
from QHyper.util import (
    weighted_avg_evaluation, sort_solver_results, add_evaluation_to_results)

problem = solver.problem

# Evaluate results with weighted average evaluation
print("Evaluation:")
print(weighted_avg_evaluation(
    results.probabilities, problem.get_score,
    penalty=0, limit_results=10, normalize=True
))
print("Sorted results:")
sorted_results = sort_solver_results(
    results.probabilities, limit_results=5)
print(sorted_results)

# Add evaluation to results
results_with_evaluation = add_evaluation_to_results(
    sorted_results, problem.get_score)

for rec in results_with_evaluation:
    print(f"Result: {rec}, "
          f"Prob: {rec['probability']:.5}, "
          f"Evaluation: {rec['evaluation']:.5}")

Evaluation:
-1.669217721264391

Sorted results:
[(1, 1, 0, 0, 1, 0.14605589) (1, 0, 1, 0, 1, 0.09231208)
 (0, 1, 1, 0, 1, 0.09231208) (1, 0, 1, 1, 0, 0.06831021)
 (0, 1, 1, 1, 0, 0.06831021)]
 
Result: (1, 1, 0, 0, 1, 0.14605589, -4.), Prob: 0.14606, Evaluation: -4.0
Result: (1, 0, 1, 0, 1, 0.09231208, -3.), Prob: 0.092312, Evaluation: -3.0
Result: (0, 1, 1, 0, 1, 0.09231208, -3.), Prob: 0.092312, Evaluation: -3.0
Result: (1, 0, 1, 1, 0, 0.06831021, 0.), Prob: 0.06831, Evaluation: 0.0
Result: (0, 1, 1, 1, 0, 0.06831021, 0.), Prob: 0.06831, Evaluation: 0.0

\end{lstlisting}
QHyper provides a set of Jupyter Notebooks with simple and real-life examples of solving combinatorial optimization problems.
\section{Impact}
As was presented in this paper, QHyper is a Python library 
designed for a convenient execution of hybrid quantum-classical combinatorial optimization experiments. Ranging from providing a simple and extensible interface for problem formulation, through a variety of solvers (quantum gate-based, annealers, and classical ones), to (hyper)parameter optimization, the library can be utilized in diverse scenarios. The simplest use of QHyper is as a stand-alone tool for solving various optimisation problems. \\

What is more, QHyper improves users' way of testing the advancements of their computational experiments. One can easily implement and extend the library with well-known solver metrics (e.g., solution quality, solution diversity,  time-to-solution, or cost-to-solution~\cite{abbas_quantum_2023}). Then, QHyper can be used for quick testing in order to find the best trade-off between speed and quality of results.  
As such, it is an excellent tool for non-experts lowering their barrier of learning new field. \\

QHyper's main strength lies in its unifying approach. Users can easily interconnect their scientific problem solving software to QHyper and test various solvers for research results. QHyper provides also (hyper)parameter optimization algorithms for existing solvers which can be easily applied in cases where hyperparameters are necessary to properly express the problem formulation or where fine-tuning parameters is a significant aspect of the solver's operation, as is the case in VQAs. \\

Together with simple configuration options and easy management via Jupyter Notebooks, QHyper has the potential to be widely applied by practitioners, engineers or academics pursuing their research and to become the first choice library for supporting combinatorial optimization. In fact, QHyper is currently being used for optimization of workflow scheduling~\cite{krzhizhanovskaya_foundations_2020}as well community detection in brain connectomes~\cite{wierzbinski_community_2023}.


  
\section{Summary and Conclusions}
Any advancement brought by quantum (or hybrid quantum-classical) algorithms over classical ones can give a significant impact on the high number of optimization problems that arise in many science domains. Although there exist hybrid quantum-classical and quantum computing optimization software libraries, most of them specialize exclusively on particular topics and lack general approach. We have therefore proposed the QHyper platform which changes this angle and thus address users' needs concerning various combinatorial optimization tasks. The library provides unifying architecture covering diverse requirements of computational experiments ranging from formulating  combinatorial problems, through selecting and running chosen solvers until objective function hyperparameters optimisation. 
The rich set of solvers and 
%
the designed interface allows users to add custom optimization problems, own solvers, and necessary optimizers making QHyper extensible on every level of usability.\\

In the future, in  the case of solvers based on variational algorithms, we plan to add connectivity to real gate-based hardware backends. Additionally, there is a need to develop best practices for  researchers using both QHyper and  well-established domain-specific classical solvers to facilitate their investigation, if quantum optimization methods are promising for their field. 

\section*{Acknowledgements}
\label{}
\textit{The research presented in this paper received support from the funds assigned by Polish Ministry of Science and Technology to AGH University. We gratefully acknowledge Polish high-performance computing infrastructure PLGrid (HPC Center: ACK Cyfronet AGH) for providing computer facilities and support within computational grant no. PLG/2024/017208
}



\bibliographystyle{elsarticle-num} 
\bibliography{referencesKR}







\end{document}